\documentclass[a4paper,twocolumn,accepted=2021-07-22]{quantumarticle}
\pdfoutput=1

\usepackage[english]{babel}
\usepackage{graphicx}
\usepackage{amsmath,amssymb,amsfonts, amsthm}
\usepackage[breaklinks=true]{hyperref}

\usepackage[square,numbers,compress]{natbib}
\usepackage[T1]{fontenc}
\usepackage{physics} 

\usepackage{comment} 

\usepackage{tikzit}

\tikzstyle{trapezium}=[fill=white, draw=black, shape=trapezium, trapezium stretches=true, trapezium right angle=-75, trapezium left angle=0, minimum width=0.75cm, minimum height=0.5cm, line width=0.025cm]
\tikzstyle{bigtrapezium}=[fill=white, draw=black, shape=trapezium, trapezium stretches=true, trapezium right angle=-75, trapezium left angle=0, minimum width=1.1cm, line width=0.025cm]
\tikzstyle{BIGtrapezium}=[fill=white, draw=black, shape=trapezium, trapezium stretches=true, trapezium right angle=-75, trapezium left angle=0, minimum width=1.5cm, line width=0.025cm]
\tikzstyle{daggerbigtrapezium}=[fill=white, draw=black, shape=trapezium, trapezium stretches=true, trapezium right angle=75, trapezium left angle=0, minimum width=1.1cm, line width=0.025cm]
\tikzstyle{state}=[fill=white, draw=black, shape=isosceles triangle, isosceles triangle stretches=false, inner sep=0, shape border rotate=-90, isosceles triangle apex angle=75, minimum width=0.6cm, line width=0.025cm]
\tikzstyle{effect}=[fill=white, draw=black, shape=isosceles triangle, isosceles triangle stretches=false, inner sep=0, shape border rotate=90, isosceles triangle apex angle=75, minimum width=0.6cm, line width=0.025cm]
\tikzstyle{discard}=[shape=tlground, fill=white, draw=black, rotate=180]
\tikzstyle{identity}=[shape=tlground, fill=white, draw=black]
\tikzstyle{SAdiscard}=[shape=ground, fill=white, draw=black, rotate=180]
\tikzstyle{SAidentity}=[shape=ground, fill=white, draw=black]

\tikzstyle{black}=[-, draw=black, line width=0.025cm]

\usepackage[]{circuitikzgit}

\usepackage{diagrams}
\newarrow{Equiv}{<}{-}{-}{-}{>}

\usepackage{xspace}

\newcommand{\foottext}[1]{\footnotetext[\value{footnote}]{#1}}

\newcommand{\ie}{{\textit{i.e.}}\xspace}

\newcommand{\id}{\mathbb{I}\xspace}

\newtheorem{mydef}{Definition}

\usepackage[autostyle, english = american]{csquotes}
\MakeOuterQuote{"}

\usepackage[normalem]{ulem}

\begin{document}

\title{The arrow of time \newline in operational formulations of quantum theory}

\author{Andrea Di Biagio}%
\email{andrea.dibiagio@uniroma1.it}%
\affiliation{Dipartimento di Fisica, Sapienza Universit\`a di Roma, 00185 Roma, Italy}%
\orcid{0000-0001-9646-8457}%
\author{Pietro Don\` a}%
\email{pietro.dona@cpt.univ-mrs.fr}%
\affiliation{Aix-Marseille Universit\'e, Universit\'e de Toulon, CNRS, CPT, 13288 Marseille, France}%
\orcid{0000-0001-7341-0682}%
\author{Carlo Rovelli}%
\email{rovelli@cpt.univ-mrs.fr}%
\affiliation{Aix-Marseille Universit\'e, Universit\'e de Toulon, CNRS, CPT, 13288 Marseille, France}%
\affiliation{Perimeter Institute, 31 Caroline Street North, Waterloo Ontario N2L2Y5, Canada}%
\affiliation{The Rotman Institute of Philosophy, 1151 Richmond St.\,N, 
London Ontario N6A5B7, Canada}%
\orcid{0000-0003-1724-9737}%

\begin{abstract}\noindent
The operational formulations of quantum theory are drastically time oriented. However, to the best of our knowledge, microscopic physics is time-symmetric.  We address this tension by showing that the asymmetry of the operational formulations does not reflect a fundamental time-orientation of physics. Instead, it stems from built-in assumptions about the \textit{users} of the theory. In particular, these formalisms are designed for predicting the future based on information about the past, and the main mathematical objects contain implicit assumption about the past, but not about the future. The main asymmetry in quantum theory is the difference between knowns and unknowns.
\end{abstract}

\maketitle 

\section*{Introduction}
\label{sec:time-oriented}

Classical mechanics is invariant under time reversal: its elementary laws do not distinguish past from future. The observed arrow of time is a macroscopic phenomenon that depends on the use of macroscopic variables and the contingent fact that the entropy defined by these variables was lower in the past. Is it the same for quantum mechanics?

On the one hand, the Schr\"odinger equation is time reversal invariant and so is quantum field theory (up to parity transformation and charge conjugation). Elementary physics is time reversal invariant and the source of time orientation is again  macroscopic and entropic. Elementary quantum phenomena do not carry a preferred arrow of time. On the other hand, however, the formalism of quantum theory is often defined in a markedly time oriented way.  

Here we address this tension between the physics and the formalism. We investigate the reason for the time orientation of the quantum formalism and show that the tension can be resolved. 
The asymmetry in the formalism is due to the inherent directionality in the process of inference, which is related to the arrow of time only indirectly.

Let us start by noting that, in any inferential problem, there is an asymmetry between what is known (the data), and what is unknown (the desiderata). Let us call this directionality the \textit{arrow of inference}.  The arrow of inference is not necessarily aligned with the entropic arrow of time. The arrow of inference may be pointed towards the past as well as towards the future. Quantum phenomena are such that we can only compute conditional probabilities, so quantum theory inherits the asymmetry between data and desiderata.

Quantum theory allows us to compute the probability of future events from past ones, but it also allows us to compute the probability of past events from future ones. As we illustrate in detail below, quantum theory does not distinguish between these two tasks. 
In contrast, the users of quantum theory are generally more interested in predicting the future than postdicting the past. We live in thermodynamically oriented world that has abundant macroscopic traces of the past but not of the future \cite{price1997time,rovelli2020memory}. Hence in most problems, the arrow of inference points in the same direction as the arrow of time. As a result we have designed formulations of quantum theory that conflate the two arrows. Ignoring the distinction may be a source of confusion.

We will focus on formalisms used in quantum information \cite{Mike&Ike,dariano2017quantum,PQP}. These are designed to study information processing tasks and the correlations that agents can achieve by sharing and manipulating quantum systems. This approach has lead to a wealth of insights, both of theoretical and technological value  \cite{bell1964einstein,bell1966problem, wootters1982single,shor1997polynomial,ekert1991quantum,grover1996fast,bennett2014quantum,dieks1982communication,bennett1993teleporting, holevo1996capacity,chiribella2013quantum,giacomini2019quantum,bose2017spin,marletto2017gravitationallyinduced,bong2020strong}.
In particular, the information-theoretic reconstructions of quantum theory \cite{hardy2001quantum,dakic2009quantum,masanes2011derivation,chiribella2011informationala,hardy2013reconstructing,hoehn2017toolbox,hoehn2017quantuma,selby2018reconstructing,jia2018quantum,oeckl2019local} derive the formal Hilbert space structure of quantum theory from simple physical principles (much like Einstein's two postulates of special relativity allowed to re-derive the Lorentz transformations \cite{rovelli1996relational,fuchsQuantumMechanicsQuantum2002}). With the notable exception of Ding Jia's \cite{jia2018quantum}, the reconstructions either start by considering a space of theories that is intrinsically time-oriented \cite{hardy2001quantum,dakic2009quantum,masanes2011derivation,hardy2013reconstructing,hoehn2017toolbox,hoehn2017quantuma,jia2018quantum} or introduce the time orientation explicitly as a postulate \cite{chiribella2011informationala,selby2018reconstructing,oeckl2019local} ("no signaling from the future"). 
 
There is nothing wrong with time-oriented formalisms designed to study time-oriented questions. But the success of this approach risks obscuring the time reversal invariance of elementary physics, and this could turn into an obstacle when extending it to the investigation of phenomena outside the laboratory.
An example of such situations is quantum gravitational phenomena, in which the existence of a background spacetime cannot be taken for granted \cite{qiss, dibiagio2020can}. 

During the development of this work, Schmid, Selby, and Spekkens \cite{schmid2020unscrambling}, and Hardy \cite{hardy2021time} have proposed formalisms in which the physical and the inferential aspects of a theory can be separated and make space for time-symmetric physics. The present work can be seen as an additional motivation for these frameworks.

Our argument proceeds as follows. We start with the uncontroversial assumption that the Born rule gives \emph{prediction}{ probabilities}: conditional probabilities for future events, given past ones. We apply standard probability theory to find formulas for \emph{postdiction}{ probabilities}: probabilities about the past, given the future. Note that we do not postulate these probabilities: we derive them from the prediction probabilities.

In section \ref{sec:unitary}, we show that for closed quantum systems the probabilities for prediction and postdiction are identical, a property we call \textit{inference symmetry}. The Born rule can be used in both directions of time without modification, contrary to what is sometimes stated. 

In section \ref{sec:open}, we discuss open quantum systems, where this symmetry is hidden. In that case, the prediction and postdiction probabilities differ. However the difference is dictated by the asymmetries in the inferential problem, not by the arrow of time, once again contrarily to what is often stated in the literature. Unitary quantum mechanics is both time-symmetric and inference-symmetric. 

In section \ref{sec:quantum-channels}, we investigate the same question for \textit{quantum channels}, the more general evolutions featuring in operational formulations of quantum theory used in quantum information. Quantum channels are not, in general, inference-symmetric.  By shifting the Heisenberg cut to include part of the apparatus, we show that the inference-asymmetry of quantum channels stems from asymmetries in the inferential data.

In section \ref{sec:timereversal}, we relate the tasks of postdiction with passive and active time-reversals, and discover that quantum channels can also be seen as shorthands for calculations about the past.

We combine all insights in section \ref{sec:causality} to show that the asymmetries of the quantum information  formulations do not stem from an arrow of time intrinsic to all quantum systems, but from the asymmetry inherent in the process of inference. The time-asymmetry of the operational formalisms used in quantum information theory is that of the time-oriented macroscopic agents that set up the experiments.

 At the end of the paper, we briefly discuss the time orientation of other formulations (Copenhagen, Everettian, de Broglie-Bohm) as well.

\section*{Previous work}

The exploration of the time-symmetry of quantum uncertainty started early with Einstein, Tolman, and Podolski \cite{einstein1931knowledge} already noting in 1931 that ``the principles of quantum mechanics actually involve an uncertainty in the description of past events which is analogous to the uncertainty in the prediction of future events.''

Aharonov, Bergmann, and Lebowitz \cite{aharonov1964time} build a time-symmetric theory out of quantum theory by considering the frequencies of outcomes of a sequence of projective measurements in ensembles constructed by pre- and post-selection. They note that this theory is time-symmetric in the sense that the frequencies observed in one ensemble are the same as the ones observed in the ensemble prepared by swapping the pre- and post-selection and performing the measurements in the reversed order. They note that probabilities calculated based only on preselection are experimentally accurate, while probabilities calculated only on post-selection are insufficient. They argue that this time-asymmetry is not inherent to quantum mechanics but is a consequence of the asymmetry of the macroscopic world. Finally they advance that this asymmetry should be represented by adding a time-asymmetric postulate to the time-symmetric theory. In light of our work, we don't need to build a time-symmetric theory and then break the symmetry. Quantum mechanics is already symmetric in the sense that generally it does not distinguish prediction and postdiction, or the past from the future. Our work also has a different aim: to understand the time-asymmetry of the operational formulations of quantum theory.

While Aharonov \textit{et. al.} \cite{aharonov1964time} considered a sequence of measurements ``sandwiched'' between selection events, the setup in our work is similar to that in \cite{watanabe1955symmetry}, where Watanabe was concerned with calculating the probabilities for a past or future event given a present event. Watanabe introduced \textit{retrodictive quantum mechanics}, where a state is assigned to the system based on present data and then evolved to past times. Like Aharonov \textit{et. al.}, Watanabe remarks that ``blind retrodiction,'' (\textit{i.e.} what we call postdiction: retrodiction with flat priors on the past events) does not work well in practice because agents in the past can decide to interrupt or go on with the experiments. Watanabe also recognises that inference is inherently asymmetric and can run in either direction of time, starting from data at a given instant.

Retrodictive quantum mechanics has been further developed over the decades by Barnett, Pegg, Jeffers and collaborators \cite{barnett2000bayes,pegg2002quantum,speirits2017retrodiction} who devise general postdiction formulas, including ones equivalent those we derive in sections \ref{sec:unitary} and \ref{sec:pre-post-channels} for postdiction in closed systems and for quantum channels respectively. Our contribution to retrodictive quantum mechanics is twofold. First, we propose the explanation of the asymmetry between prediction and postdiction in quantum channels in terms of implicit data about the past of a purifying system.
Secondly, we prove \textit{no-signalling from the unknown}, which is a property of quantum mechanics, and in particular of retrodictive quantum mechanics, that to our knowledge has not been recognised before.
We also pay attention to the conceptual difference between retrodiction and time-reversal, and we relate these two concepts.

Leifer and Pusey \cite{leifer2017time} consider a similar prepare-and-measure scenario as us and investigate what time-symmetry can imply on possible ontological models for quantum theory. Their definition of an \textit{operational time reverse} formally equivalent to what we call an active time-reversal. They define \textit{operational time-symmetry} as the existence of an operational time reverse. They derive a fascinating no-go theorem for  ontic extensions (aka hidden variable models) of operational time-symmetric processes. We share the belief that operational time-symmetry is an essential feature of quantum mechanics, but we do not concern ourselves with ontic extensions. Instead, we are interested in understanding why not all quantum channels are operationally time-symmetric. We also study the difference and relation between postdiction and time-reversal,  prove that operational time-symmetry is equivalent to inference symmetry.

Oreshkov and Cerf \cite{oreshkov2015operational} define an extension to operational quantum theory, allowing for a ``notion of operation that permits realizations via both pre- and post-selection.'' Their motivation for building a new theory is stated in the abstract: ``The symmetry of quantum theory under time reversal has long been a subject of controversy because the transition probabilities given by Born's rule do not apply backward in time.'' Our work shows that the Born rule applies equally well in both directions in time---as long as we treat prediction and postdiction on equal footing. We argue that the asymmetry of operational quantum theory reflects the asymmetry of the agents, not the asymmetry of quantum phenomena per se.

Our work was inspired by conversations at the QISS conference at HKU and at the QISS virtual seminars \cite{oeckl2020what}, where it transpired that the time-asymmetric operational formulation obfuscated the fundamental time-symmetry of quantum theory.

The importance of separating the physical from the inferential in quantum mechanics is a more modern idea, perhaps traceable to E.T. Jaynes who  famously compared quantum theory to an omelette to be unscrambled \cite{jaynes1990probability}. The QBists see quantum theory as not much more than the correct probability calculus to use in our world \cite{Fuchs2019}. Leifer and Spekkens \cite{leifer2013formulation} have formally developed the analogy between quantum probabilities and Bayesian inference. They introduce a notion of ``quantum conditional states'' representing sequential and parallel quantum experiments and prediction and postdiction on the same footing, just like in classical probability theory. In our work, we limit ourselves to classical probability theory. At first, we use the Born rule to obtain classical conditional probability distributions $P_{pre}(x|a,U)$ for prediction probabilities in sequential experiments. We show that the Born rule actually can be used to compute prediction and postdiction probabilities.

\medskip
We summarize the main original contributions of this work.
\begin{itemize} \setlength\itemsep{0em}
    \item We interpret the inference-asymmetry of quantum channels as due to implicit information about the purifying system. (Section \ref{sec:purification}).
    \item We note that the asymmetry of the operational formulation of quantum mechanics is due to three factors. 
    \begin{enumerate}
        \item The asymmetry intrinsic in the process of inference (Section \ref{sec:causality})
        \item The operational formulation is interested in prediction.(Section \ref{sec:operations})
        \item  Quantum operations contain built-in assumptions about the past and the ability of agents to pre-select states, and this is where the time-symmetry is broken. (Sections \ref{sec:purification} and \ref{sec:agents})
    \end{enumerate}
    \item The proof of the \textit{no-signalling from the further unknown} property of quantum mechanics, a consequence of the conservation of probability and not an independent axiom. (Section \ref{sec:further-unknown}) When applied to prediction, it reduces to the well-known ``no-signalling from the future.'' 
    \item The derivation that inference-symmetry is equivalent to bistochasticity (Section \ref{sec:inference-symmetry}) and hence to operational time-symmetry (Section \ref{sec:time-reversed-quantum-channel}). 
\end{itemize}

\section{Prediction and postdiction}
\label{sec:pre-post}

Quantum indeterminism is time-reversal invariant. In presenting the probabilistic nature of quantum theory, we often emphasise that the future of a quantum system is not entirely determined by its past. It certainly is true that the outcomes of future interactions with a quantum system are uncertain, given the details of past interactions. What is less often recognised is that the converse is equally true: given the details of present interactions, the past ones are uncertain. This was already pointed out a long time ago \cite{einstein1931knowledge}: the irreducible indeterminism of quantum phenomena cuts both ways, leaving both the past and the future uncertain, given data about the present. This has practical consequences such as the impossibility of deterministic state-discrimination and the no-cloning theorem \cite{Mike&Ike}: interactions with a quantum system do not allow us to guess with certainty how it was prepared.

As we will see below, the past of a quantum system is \textit{quantitatively} as uncertain as its future: the probabilities calculated using the Born rule can be applied to both predict and postdict. Quantum theory, in fact, does not distinguish \textit{a priori} the tasks of prediction and postdiction and we might say that there is a fundamental "unpostdictability" of the behaviour of quanta. 

Let us operationally define what we mean by prediction and postdiction using two related tasks.
In both tasks, a friend prepares a quantum system in an initial configuration, allows it to undergo a given transformation, measures it, and finds it in some final configuration. The friend then gives us some information and about these events, and asks us to guess the rest.
In the first task, we are asked to guess the outcome of the measurement, given the initial configuration and the details of the transformation and of the measurement. In the second task, we guess the outcome of the preparation, given the outcome of the measurement, the details of the transformation, and the set of possible initial states.

\begin{mydef}[Prediction task]
Given a preparation, a map and a test and the outcome of the preparation, calculate the probabilities for the outcomes of the test.
\end{mydef}
\begin{mydef}[Postdiction task]
Given a preparation, a map and a test and the outcome of the test, calculate the probabilities for the outcomes of the preparation.
\end{mydef}
These are inferential tasks, in which we use the available information to make educated guesses.
If we want to think of probabilities in frequentist terms, we can imagine the friend repeating the setup many times, ensuring a uniform distribution in the initial or final configuration and interpret the calculated probabilities as relative frequencies of outcomes in an ensemble of trials, in the limit of infinite trials. 

While the setting of these tasks might appear artificial at first, a moment of thought reveals that it serves as a useful shorthand for physically relevant situations.
In fact, postdiction has been extensively studied before \cite{barnett2000bayes,barnett2014quantum,pegg2002quantum,fields2020quantum} and has a number of practical applications; see \cite{speirits2017retrodiction,yuan2018experimental,rossi2019observing} and references therein. 

In the following, we study the relation between these two tasks, as it captures the role played by the arrow of time in quantum theory. 
First, we consider the case of closed systems,  namely when the system under consideration is isolated between preparation and observation. Then we consider open systems, namely when we ignore some degrees of freedom, such as environmental degrees of freedom.  Finally, we extend the analysis to the more general case in which the notions of preparation, evolution and measurement are subsumed in the more general idea of \emph{operation} used in quantum information and quantum foundations.

\section{Closed systems}
\label{sec:unitary}
In this section, we assume that the friend prepares the system by determining the values of a maximal set of compatible observables and does the same at observation. We also assume that the system under consideration is isolated between preparation and observation. Therefore, the preparation and test are represented by orthonormal bases of the Hilbert space associated with the quantum system and the transformation is represented by a unitary transformation.

In this case, the Born rule is equally good for predicting the future given the past and postdicting the past given the future \cite{watanabe1955symmetry}. Since this fact is not universally known, we derive it assuming only the uncontroversial fact that the Born rule can be used to predict the future.

We denote by $\{a_i\}_{i=1}^d$ and $\{x_i\}_{i=1}^d$ the bases of the preparation and test, respectively---although we drop the basis indices when they are not strictly needed, to keep the notation cleaner. 
The solution to the prediction task with the unitary evolution $U$, and the outcome of the preparation $a$ is given by the Born rule: the probability 
\begin{equation}\label{bornU}
P_{pre}(x|a,U) = |\!\matrixel{x}{U}{a}\!|^2 ,
\end{equation}
for the outcome $x$ of the test. 
The solution to the postdiction task is obtained from the solution of the prediction game and standard probability theory. By Bayes' theorem
\begin{equation}\label{BayesPostU}
P_{post}(a|x,U) = \frac{P_{pre}(x|a,U)P(a)}{P(x)},
\end{equation}
where $P(a)$ is the prior probability on the initial configuration and $P(x)$ is the probability of the final configuration given the prior.
Since all we know in the postdiction task is the basis of the preparation, we have
\begin{equation}\label{prior}
P(a_i) = \frac{1}{d}
\end{equation}
for all $i=1,\dots d$.
The \textit{a priori} probability of the outcome of the test is computed summing over all possible initial states:
\begin{equation}
P(x) = \sum_{i=1}^{d} \frac{1}{d}P_{pre}(x|a_i,U)=\frac{1}{d} ,
\end{equation}
where we used the fact that the evolution is unitary. The postdiction probability is then computed from \eqref{BayesPostU}:
\begin{equation}\label{postU}
P_{post}(a|x,U) = |\matrixel{x}{U}{a}|^2 = P_{pre}(x|a,U),
\end{equation}
which, in terms of the pictorial calculus of Ref. \cite{PQP}, reads
\begin{equation}
P_{pre}(x|a,U) = \scalebox{1.25}{\begin{tikzpicture}
	\begin{pgfonlayer}{nodelayer}
		\node [style=trapezium] (0) at (0, 0) {$U$};
		\node [style=state] (1) at (0, -1.5) {$a$};
		\node [style=effect] (2) at (0, 1.5) {$x$};
	\end{pgfonlayer}
	\begin{pgfonlayer}{edgelayer}
		\draw [style=black] (2) to (0);
		\draw [style=black] (0) to (1);
	\end{pgfonlayer}
\end{tikzpicture}
} = P_{post}(a|x,U).
\end{equation}
Thus for a closed quantum system, the solution to the prediction and postdiction tasks is given by the same formula.

Note that the flat prior \eqref{prior} is crucial in the derivation above. Had the prior been different, the postdiction probabilities would be different from the prediction probabilities. However, assuming a different prior would introduce an inappropriate asymmetry between the two tasks. Our objective is to treat the prediction and postdiction tasks on equal footing. When we predict the result of a measurement using the Born rule, we do not assume any prior knowledge on the result of the experiment besides the space of alternatives. Therefore, in a postdiction task we do not assume any prior knowledge of the result of the preparation besides the space of alternatives, hence the the flat prior. The flat prior does not imply that the input system was prepared in the maximally mixed state, it is simply the probability distribution that represents the prior knowledge in the postdiction task.

For a system evolving unitarily between the preparation and the observation events,
later events are uncertain given the earlier event \textit{and} earlier events are uncertain given later events, and the probabilities are given by the same formula. The Born rule does not distinguish the past from the future: it allows to calculate the probability of an event given another event, no matter their order in time.

Let us formalise this property:
\begin{mydef}[Inference symmetry] 
A transformation $\Phi$ is inference symmetric if for any two orthonormal bases $\{a_i\}_{i=1}^{d_A}$ and $\{x_i\}_{i=1}^{d_X}$ for the input and output spaces respectively, the prediction and postdiction probabilities are identical for given $i$ and $j$:
\begin{equation}\label{postdictionsymmetry}
P_{pre}(x_i|a_j,\Phi) = P_{post}(a_j|x_i,\Phi)
\end{equation}
\end{mydef}
\noindent Quantum unitary evolution is inference-symmetric.
Why is this time-symmetric aspect of unitary evolution rarely emphasised?

In most practical situations, we don't need to use the flat prior when guessing the past. We can do better, because there are macroscopic traces of the past, like our memories or entries in a notebook.  Additionally, in most laboratory experiments, the initial distribution is known or chosen so that frequency of events in the ensemble is far from uniform. 

The fact that macroscopic traces are records of the past and not the future and the fact that an experimenter's choice can affect the future and not the past are both macroscopic phenomena that pertain to the irreversible physics of the macroscopic world surrounding the experiment \cite{rovelli2020memory,rovelli2020agency}, not to the quantum dynamics, which by itself does not know the arrow of time.

Some authors go so far as to say that the Born rule does not work `backward in time' and see it as a fundamental asymmetry in the theory \cite{oreshkov2015operational} and that quantum theory needs to be modified, or extended, to make it symmetric. But this is too quick.  If we do not assume any knowledge or bias in the past, \eqref{postU} is indeed the correct formula to use according to quantum theory. In turn, the validity of the Born rule in predicting the future relies on the same assumptions about the future. Namely, if we did have some knowledge of the future, then \eqref{bornU} would not be the best formula to make predictions. 
For example, if the detector does not detect certain states, the Born rule fails.

The presence of records of the past is not a property of quantum theory \textit{per se}, or the behaviour of a single quantum, but a property of what surrounds the quantum. See also \cite{watanabe1955symmetry, aharonov1964time} for early examples of this argument.

\section{Open systems}
\label{sec:open}

Let us now consider the case when the system we deal with is not isolated.  An open quantum system can always be seen as a part of a larger closed quantum system.  To study this case, consider a tensor decomposition of the input $A\otimes B$ and output $X\otimes Y$ Hilbert spaces.  The tasks we consider now regard computing probabilities restricted to some of these subspaces.  Denote by $d_A$, $d_B$, $d_X$ and $d_Y$ the dimensions of the respective spaces and with $\{a_i\}$, $\{b_i\}$,  $\{x_i\}$, and $\{y_i\}$ bases on them. 
The evolution between input and output space is represented by the unitary $U$.
By the results of the previous subsection, this process is inference-symmetric with the solution
\begin{equation}
\label{symmetryPart}
    P_{pre}(xy|ab,U) = |\matrixel{xy}{U}{ab}|^2 = P_{post}(ab|xy,U).
\end{equation}

Suppose that we agree with our friend to ignore the subspace $Y$ of the outcome space and compute only the probability of finding $x$ as the outcome of the test on $X$. This simulates the situation in which our system gets entangled with some other system that is subsequently ignored, like when information leaks into the environment. Note that the difference between a closed and an open system is in the inferential data, not in the physical system. We can solve the prediction task by computing the marginal probability
\begin{equation}\label{predictionPart}
	P_{pre}(x|ab,U) = \sum_{i=1}^{d_Y}P_{pre}(xy_i|ab,U),
\end{equation}
where we sum over the space $Y$ we decided to neglect. Similarly we solve the postdiction task with the same unitary evolution, with only knowledge on the outcome of the test on the space $X$, weighting the original postdiction probabilities with a flat prior:
\begin{equation}\label{postdictionPart}
	P_{post}(ab|x,U) = \sum_{i=1}^{d_Y}\frac{1}{d_Y}P_{post}(ab|xy_i,U).
\end{equation}
We can use the inference symmetry of the closed system \eqref{symmetryPart} to relate the two expressions above:
\begin{equation}\label{post-missing-output}
	P_{post}(ab|x,U) = \frac{1}{d_Y} P_{pre}(x|ab,U).
\end{equation}
Thus the prediction and postdiction probabilities are no longer equal once part of the output system is ignored.

Suppose now that we agree with our friend to neglect the $B$ part of the input space. This corresponds to the situation in which a system in an unknown state interacts with our original system. Again, the difference is in the inferential data, not the physical setup. The prediction probabilities are obtained by assigning a flat prior to the system $B$:
\begin{equation}
\begin{aligned}\label{pre-missing-input}
	P_{pre}(xy|a,U)
	&= \sum_{i=1}^{d_B}\frac{1}{d_B}P_{pre}(xy|ab_i,U),
\end{aligned}
\end{equation}
while the postdiction probabilities are
\begin{equation}
	P_{post}(a|xy,U) = \sum_{i=1}^{d_B}P_{post}(ab_i|xy,U).
\end{equation}
The two are again related using \eqref{symmetryPart}: 
\begin{equation}\label{post-missing-input}
	P_{post}(a|xy,U) = d_B P_{pre}(xy|a,U).
\end{equation}
Again, the two probabilities are different.

We can similarly analyse the case where we agree with our friend to neglect the result of the preparation in $B$ and the outcome of the test on $Y$, which simulates a situation in which the system is open to influences from an unobserved quantum system. The prediction and postdiction probabilities again differ by a simple normalisation constant: 
\begin{equation}\label{post-missing-both}
	d_Y P_{post}(a|x,U) = d_B P_{pre}(x|a,U).
\end{equation}
Note that the probabilities are equal only when $d_B=d_Y$, so that $A\equiv X$. When the input and output spaces are treated symmetrically, prediction and postdiction tasks are symmetric.

Crucially, the normalisation factor that makes the two kinds of probabilities different does not depend on time: if we neglect a subsystem $B$ at time $t_1$ and another system $Y$ at time $t_2$, then the probabilities for the values at $t_2$ given the values at $t_1$ are $d_Y/d_B$ times the probabilities of guessing the values at $t_1$ given the values at $t_2$. This has nothing to do with a pre-established direction of time. Indeed, it is true regardless of whether $t_1<t_2$ (as in the example above) or $t_1>t_2$.

In the general case, the asymmetry between the prediction and postdiction tasks arises because of an asymmetry in the inferential data, not because of an intrinsic asymmetry in the physics or the evolution of the system. Indeed, in all these cases, the underlying process $U$ is inference-symmetric. The inferential tasks are asymmetric only when the inferential data are asymmetric.

\subsection*{The normalisation of the identity is determined by the arrow of inference}
\label{sec:normalisation}
We can rephrase all the probability calculations above in terms of density operators. Under unitary evolution a density operator transforms as $\rho\mapsto U[\rho]:=U\rho U^\dagger$. A pure state $\psi$ can be represented as a density operator by the projector $\ketbra{\psi}$ and the Born rule can be recast as a trace 
\begin{equation}\label{bornpartial}
|\!\matrixel{xy}{U}{ab}\!|^2=\tr\big(\ketbra{xy}\; U[\ketbra{ab}] \big).
\end{equation}
Let us start with prediction, the more familiar task. In this language the prediction probability \eqref{predictionPart} can be rewritten as
\begin{equation}
\begin{aligned}
    P_{pre}(x|ab,U)
    &= \sum_{i=1}^{d_Y}P_{pre}(xy_i|ab,U) \\
    &= \sum_{i=1}^{d_Y}\tr\big(\ketbra{xy_i} \; U[\ketbra{ab}]\big) \\
    P_{pre}(x|ab,U)
    &= \tr\big((\ketbra{x} \otimes \id_Y) \; U[\ketbra{ab}]\big) .
\end{aligned}
\end{equation}
The decision to ignore part of the output system is represented by the insertion of the identity operator $\id_Y$, which in this role is called the discard operator of the subsystem $Y$.
This is the classic technique of `tracing out' a subsystem to ignore its future.
Equation \eqref{pre-missing-input} can be similarly recast as
\begin{equation}
    P_{pre}(xy|a,U) =\tr\left(\ketbra{xy} \; U\left[\ketbra{a}\otimes \frac{1}{d_B}\id_B\right]\right),
\end{equation}
where the flat prior is represented by the maximally mixed state, the density operator $\id_B/d_B$. 
This is also the well known result of doing prediction based on partial information.
Note that the two formulas above make it clear that the choice of basis of the ignored system is irrelevant to the computed probabilities.

Let us now look at the postdiction probabilities. Using the two equations above together with \eqref{post-missing-output} and \eqref{post-missing-input}, we can immediately write
\begin{align}
P_{post}(ab|x,U) &= \tr\left(\!\left(\ketbra{x} \otimes \frac1{d_Y}\id_Y\right) U[\ketbra{ab}]\right), \\
P_{post}(a|xy,U) &= \tr\big(\ketbra{xy} \; U[\ketbra{a}  \otimes \id_B]\big).
\end{align}
We see that the identity operator appears again in the ignored systems. However, the normalisation is the opposite of the predictive case.

The discard operator and the maximally mixed state are well known, and are normally only applied to the output and input side respectively. But
the normalisation of the identity operator does not reflect the direction of time, the past or the future, input or output. It reflects the direction of inference. This is particularly obvious rewriting  in the  pictorial calculus:
\begin{align}
  \scalebox{1}{$P_{pre}(x|a,U)$} \!=\!\!\! 
\begin{tikzpicture}
	\begin{pgfonlayer}{nodelayer}
		\node [style=bigtrapezium] (0) at (0, 0) {$U$};
		\node [style=effect] (1) at (-0.625, 1.5) {$x$};
		\node [style=state] (4) at (-0.625, -1.5) {$a$};
		\node [style=discard] (5) at (0.625, 1.275) {};
		\node [style=identity, label={right:$\;\frac{1}{\,d_{\!B}}$}] (6) at (0.625, -1.28) {};
	\end{pgfonlayer}
	\begin{pgfonlayer}{edgelayer}
		\draw [style=black] (1) to (4);
		\draw [style=black] (5) to (6);
	\end{pgfonlayer}
\end{tikzpicture}   
~~~\scalebox{1}{$P_{post}(a|x,U)$}\! = \!\!\!
\begin{tikzpicture}
	\begin{pgfonlayer}{nodelayer}
		\node [style=bigtrapezium] (0) at (0, 0) {$U$};
		\node [style=effect] (1) at (-0.625, 1.5) {$x$};
		\node [style=state] (4) at (-0.625, -1.5) {$a$};
		\node [style=discard, label={right:$\;\frac{1}{\,d_{Y}}$}] (5) at (0.625, 1.275) {};
		\node [style=identity] (6) at (0.625, -1.28) {};
	\end{pgfonlayer}
	\begin{pgfonlayer}{edgelayer}
		\draw [style=black] (1) to (4);
		\draw [style=black] (5) to (6);
	\end{pgfonlayer}
\end{tikzpicture}
\end{align}
with \scalebox{0.75}{$\begin{tikzpicture}
	\begin{pgfonlayer}{nodelayer}
		\node [style=SAdiscard] (0) at (0, -0.5) {};
		\node [style=none] (1) at (0, 0) {};
	\end{pgfonlayer}
\end{tikzpicture}
$} and \scalebox{0.75}{$\begin{tikzpicture}
	\begin{pgfonlayer}{nodelayer}
		\node [style=SAidentity] (0) at (0, 0.5) {};
		\node [style=none] (1) at (0, 0) {};
	\end{pgfonlayer}
\end{tikzpicture}
$} representing the identity operator as an output and input respectively. 
We discard in the direction we guess, and we have the maximally mixed state on the side of the data.

Thus the way the inference-symmetry appears to be broken in open systems reflects an asymmetry on the inferential data that is, \textit{a priori}, independent on the direction of time.

An example illustrates why it is natural that the normalisation depends on the direction of inference and is independent on the direction of time.
Consider a system evolving with the identity, \textit{i.e.} nothing happens to it. If we are told the state $a$ of the system and we are not asked to guess anything, then all the probabilities are trivially $1=\tr\ketbra{a}$. 
Conversely, if we are only told the system is in one state out of a orthonormal basis, then the probability that it is in a given state $a$ is $\frac{1}{d}=\frac{1}{d}\tr\ketbra{a}$.
When we don't guess, all the probabilities are $1$. When we are told to guess but we have no clue, our only option is to assume a uniform distribution. This is true regardless of whether we are predicting or postdicting

We often use quantum theory to predict, which is why we generally associate the normalisation factor $1/d$ to the identity operator in the input space. In practice, we are normalising our data. If we were postdicting, we would associate the normalisation factor to the identity in the output space. The operator $\id/d$ does not represent a physical fact, but the probability distribution we use to weight the conditional probabilities.

\medskip
The results of these two sections show that the irreducible quantum uncertainty applies equally to both directions of time. Indeed, when dealing with a closed quantum system, the Born rule gives both the prediction and postdiction probabilities directly. When the prediction and postdiction probabilities differ, they do so because of an asymmetry in the inferential data.

\section{Quantum Operations}
\label{sec:quantum-channels}
The transformations considered above might seem limited in scope to researchers in quantum information and quantum foundations. In these communities, the notions of preparation, evolution and measurement are subsumed by the more general notion of \textit{operation}, which reflects their more elaborate needs:  a more coarse-grained description of quantum processes, independent on the underlying dynamics, the capacity of melding classical and quantum information processing, dealing with classical uncertainty and so on. However, since agents and labs are made of atoms and photons, and the interactions between atoms and photons is satisfyingly described by the unitary evolution and pure state approach, the results of the previous section have bearing on quantum operations too.

After a brief survey of the notion of quantum operation, we solve the prediction and postdiction tasks for a general quantum channel and explain their postdiction asymmetry.

\subsection{Operations}
\label{sec:operations}

Here we provide a short description of the notion of operation (for a more careful introduction see for example \cite{dariano2017quantum,PQP,Mike&Ike}). We then note that this notion is time-oriented by design and recall how it relates to the more basic notions.

An \textit{operation} $\mathcal{O}^{A\rightarrow X}$, also known as an \textit{instrument}, from an input Hilbert space $A$ to an output Hilbert space $X$ is represented is a set $\{O_i\} $ of completely positive (CP) trace non-increasing linear maps (aka \textit{quantum maps}) from the space $\mathcal{L}(A)$ of linear operators on $A$ to $\mathcal{L}(X)$, satisfying the \textit{completeness equation} (aka the \textit{causality condition}):
\begin{equation}\label{completeness}
  	\forall\rho\in\mathcal{L}(A):~~\sum_i \tr O_i[\rho] = \tr\rho.
\end{equation}
An operation $\mathcal{O}^{A\rightarrow X} = \{O_i\} $ also defines a completely positive, trace-preserving (CPTP) map  $\rho \mapsto \mathcal{O}[\rho]:=\sum_i O_i[\rho]$.
When the operation $\mathcal{O}^{A\rightarrow X}$ is applied to a system in state $\rho$, the outcome $i$ happens with probability given by the generalised Born rule
\begin{equation}
  	P(i|\rho,\mathcal{O}) =  \tr O_i[\rho],
 \end{equation}
and the state of the system after this outcome is
\begin{equation}\label{state-update}
  \rho_i= \frac{O_i[\rho]}{\tr O_i[\rho]}.
\end{equation}
We see that the completeness requirement \eqref{completeness} amounts to a statement of the conservation of probabilities.
If the outcome $i$ is unknown, then the state of the system is a mixture of the states above, weighed by the relevant probability:
\begin{equation}\label{unseen-operation}
  \mathcal{O}[\rho] = \sum_i P(i|\rho,\mathcal{O})\rho_i =  \sum_i O_i[\rho].
\end{equation}
If the output space of an operation $\mathcal{O}^{A\rightarrow X}$ coincides with the input space of another operation $\mathcal{M}^{X\rightarrow Y}$, these two operations can be sequentially composed forming a new operation $(\mathcal{M}\circ\mathcal{O})^{A\rightarrow Y} = \{M_j\circ O_i\}$. If the state of the input system is $\rho$, the probability of the outcome $ij$ is
\begin{equation}\label{sequential-prob}
P(ij|\rho, \mathcal{M}\circ\mathcal{O}) = \tr M_j[O_i[\rho]].
\end{equation}
Operations can also be composed in parallel using the tensor product structure of the underlying Hilbert spaces.

A \textit{preparation} of a system associated with a Hilbert space $A$ is an operation $\mathcal{P}^{I\rightarrow A}$. By a simple mathematical isomorphism, quantum maps from $I$ to $A$  can always be associated with positive linear operators on $A$, so that any preparation can be represented by a subset $\{\rho_i\}\subset\mathcal{L}(A)$ such that $\sum_i \tr\rho_i = 1$. 
Above, we have considered only preparations of the form $\{\ketbra{a_i}/{d_A}\}_{i=1}^{d_A}$, where $\{a_i\}$ is an orthonormal basis for $A$.

A \textit{test} on the same system is an operation $\mathcal{T}^{A\rightarrow I}$ from the Hilbert space $A$ to the trivial Hilbert space. By the isomorphism above, and the Riesz representation theorem, tests are often also represented by a collection of positive operators $\{\sigma_j\}\subset\mathcal{L}(A)$, such that $\sum_j\sigma_j = \id_A$ with their actions on the state given by $\rho\mapsto\tr\sigma_{j}\rho$.

An operation $\Phi^{A\rightarrow X}$ with a single outcome is also called a \textit{quantum channel}, and is represented by a CPTP map. Quantum channels are also called \textit{deterministic} quantum maps, as they have only one outcome. Above, we considered only unitary quantum channels, of the form $\rho\mapsto U\rho U^\dagger$, but much more general ones are possible. For example, consider the \textit{dephasing} channel on a qubit:
\begin{equation}
\rho\longmapsto \matrixel{0}{\rho}{0}~\ketbra{0} + \matrixel{1}{\rho}{1}~\ketbra{1}
\end{equation}
The dephasing channel can be seen as the result of applying the nondestructive projective measurement
\begin{equation}
  \big\{\rho\mapsto \ketbra{i}\rho\ketbra{i}\big\}_{i=0}^{1}
\end{equation}
and ignoring the result. 

A preparation with a single outcome is also called a \textit{state}. States are by definition represented as unit-trace positive operators, \textit{i.e.} density matrices. A test with a single outcome is also called a \textit{deterministic effect}. There is only one deterministic effect, represented by the identity operator. 

Note that this formalism is time-asymmetric \textit{by construction}. The time asymmetry shows up in two ways, reminiscent of the Copenhagen-type interpretations. First, the outcomes $\{i\}$ depend probabilistically on the state of the system in the past: the probabilities calculated in this setting are invariably \textit{prediction} probabilities. Second, the state of a system at any point in time reflects events in the past, and it is independent of the events in the future. The only data assumed to be available is data about the past.

The spaces of states and effects are not isomorphic. This was identified as the main source of time-asymmetry of operational quantum theory in \cite{oreshkov2015operational}, where it was proposed to enlarge the space of effects by not requiring that operations sum up to trace-preserving maps.
However, from the perspective of our present work, we understand this asymmetry between states and effects as being the difference between known and unknown in the process of inference. Preparations represent our assumptions in the inferential problem, while tests represent the different propositions about the unknowns. There is no need to remove the distinction between preparations and tests to make quantum theory time-symmetric, all is needed is to recognise that operational quantum theory is geared for prediction: a situation where the data is in the past of the unknowns.

\medskip
Operational quantum theory is connected to the simpler setting of pure states and unitary evolutions by the concept of \textit{purification}.
Any quantum channel $\Phi^{A \rightarrow X}$ can be purified \cite{stinespring1955positive}, meaning it can be represented by a unitary channel $U_\Phi:A\otimes B\rightarrow X\otimes Y$ and a pure state $b$ for system $B$ such that
\begin{equation}\label{stinespring}
	  \forall\rho \in\mathcal{L}(A):~\Phi[\rho] = \tr_Y U_\Phi[\rho\otimes \ketbra{b}].
\end{equation}
In other words, any quantum channel can always be understood as a unitary interaction with an ancilla quantum system prepared in a specific way and where part of the output is ignored. In fact,
any operation $\mathcal{O}^{A\rightarrow X}=\{O_i\}$ can be purified \cite{ozawa1984quantum}, meaning that it is mathematically equivalent to a unitary evolution of the system $A$ in the presence of an ancilla $B$, in which part of the output system is ignored and part of it is measured on an orthonormal basis.
That is, there exists a unitary operator $U_{\mathcal{O}}$ on the Hilbert space $A\otimes B$ and a decomposition $A\otimes B \equiv X\otimes Y\otimes Z$ and a pure state $\ketbra{b}\in\mathcal{L}(B)$ such that
\begin{equation}\label{ozawa}
\begin{split}
  \forall\rho &\in\mathcal{L}(A),\forall i:\\
  &O_i[\rho] = \tr_{YZ}\big((\ketbra{i}_Y\otimes \id_Z)\circ U_\mathcal{O}[\rho\otimes\ketbra{b}]\big),
  \end{split}
  \end{equation}
where $i$ now labels an orthonormal basis of the Hilbert space $Y$. Pictorially,

\begin{equation}
  \scalebox{1}{\begin{tikzpicture}
	\begin{pgfonlayer}{nodelayer}
		\node [style=none] (0) at (0, 0) {$=$};
		\node [style=trapezium] (1) at (-1.75, 0) {$O_i$};
		\node [style=BIGtrapezium] (2) at (2.25, 0) {$U_\mathcal{O}$};
		\node [style=none] (3) at (-1.75, -2) {};
		\node [style=none] (4) at (-1.75, 2) {};
		\node [style=none] (5) at (1.4, -2) {};
		\node [style=none] (6) at (1.4, 2) {};
		\node [style=state] (7) at (2.3, -1.5) {$b$};
		\node [style=effect] (8) at (2.3, 1.5) {$i$};
		\node [style=discard] (9) at (3.5, 1.2) {};
		\node [style=none] (10) at (3.5, 0) {};
	\end{pgfonlayer}
	\begin{pgfonlayer}{edgelayer}
		\draw [style=black] (4.center) to (3.center);
		\draw [style=black] (6.center) to (5.center);
		\draw [style=black] (8) to (7);
		\draw [style=black] (9) to (10.center);
	\end{pgfonlayer}
\end{tikzpicture}}.
\end{equation}
 
It is a well-known property of quantum theory that one can always shift the Heisenberg cut to include part of the apparatus in the quantum system under description. This is the physical content of the two mathematical results above. The label $i$ that distinguishes the various outcomes of the operation $\mathcal{O}$ is now seen as labelling the possible values of a \textit{pointer variable}, a part of the apparatus that gets entangled with the system and that, when observed, allows the determination of the state of the system.  If one cared enough, one could model every quantum operation explicitly in these terms. But countless purifications are consistent with the same operation, and these low-level details are not important when one is only interested in the effect on the given quantum systems.  Quantum operations are useful shorthand.

\subsection{Prediction and postdiction with quantum channels}
\label{sec:pre-post-channels}

We now solve the prediction and postdiction tasks using the generalised Born rule, similarly to how we did for closed quantum systems.  Unitary evolution is replaced by a CPTP map $\Phi$. The input and output spaces are $A$ and $X$ respectively, which do not need to be isomorphic. Preparation and measurement are performed on bases $\{a_i\}_{i=1}^{d_A}$ and $\{x_i\}_{i=1}^{d_X}$. The solution of the prediction task is given by the generalised Born rule: 
\begin{equation}\label{bornCPTP}
    P_{pre}(x|a,\Phi) = \tr \ketbra{x}\Phi[\ketbra{a}].
\end{equation}
The solution to the postdiction task is found again by Bayesian  inversion
\begin{equation}\label{BayesCPTP}
    P_{post}(a|x,\Phi) = \frac{P_{pre}(x|a,\Phi)P(a)}{P(x)}.
\end{equation}
Assigning a flat prior for $P(a)$, we can compute the probability of the data
\begin{equation}
    P(x) = \sum_{i=1}^{d_A} \frac{1}{d_A}P_{pre}(x|a_i,\Phi) = \tr \ketbra{x}\Phi\left[\frac{1}{d_A}\id_A \right],
\end{equation}
which this time is not uniform.
Equation \eqref{BayesCPTP} then becomes
\begin{equation}\label{postCPTP}
    P_{post}(a|x,\Phi) =
\frac{ \tr \ketbra{x}\Phi[\ketbra{a}]}
{\tr \ketbra{x}\Phi[\id_A]}.
\end{equation}
This formula was also derived in \cite{barnett2000bayes}. At first, it looks quite different from the formula for the prediction probabilities. However, it is also a remarkably simple solution. Note that the postdiction probabilities are related to the prediction probabilities by a simple multiplicative factor
\begin{equation}
P_{post}(a|x,\Phi) = f_{\Phi}(x)\cdot P_{pre}(x|a,\Phi),
\end{equation}
where
\begin{equation}
f_\Phi(x)^{-1}=\sum_{i=1}^{d_A}P_{pre}(x|a_i,\Phi) = \tr \ketbra{x}\Phi[\id_A].
\end{equation}
For a given measurement outcome $x$, the prediction probabilities are proportional to the postdiction probabilities, up to a fixed normalisation factor $f_\Phi(x)$.
Once one has calculated the set  of prediction probabilities, one already has the postdiction probabilities, up to this normalisation factor.

Inference symmetry is thus broken in general,  and this seems to support the idea there is a fundamental difference between the past and the future in quantum theory. But it is broken in a simple way, reminiscent to the situation in the previous section when we were concerned with partial data in a closed system. Indeed, when we ``look under the hood'' using purification, we see that this is essentially what is going on.
From the purified point of view, there is an asymmetry between inputs and outputs because the purifying ancilla's state is assumed known in input and ignored in output.

\subsection{Purified task}
\label{sec:purification}

Let $\Phi$ be a quantum channel from Hilbert space $A$ to Hilbert space $X$. Then there exists a unitary channel $U_\Phi:A\otimes B\rightarrow X\otimes Y$ and a pure state $b$ for system $B$ such that
\begin{equation}\label{purification}
	\Phi[\ketbra{a}] = \tr_Y U_\Phi[\ketbra{a}\otimes \ketbra{b}].
\end{equation}
The prediction probabilities for $\Phi$ are, by definition, just those of the corresponding pure open system:
\begin{align}
    P_{pre}(x|a,\Phi) =  P_{pre}(x|ab,U_\Phi),
\end{align}
as given in \eqref{predictionPart}. However, in general
\begin{align}\label{CPTP-post-neq}
    P_{post}(a|x,\Phi) \neq  P_{post}(ab|x,U_\Phi).
\end{align}
This is readily explained by the fact that the  initial state of the ancilla is assumed known, or ``held fixed.'' Since we know the ancilla's input state is $b$, the probability of the data is in fact
\begin{equation}
    P(x) = P_{pre}(x|b,U_\Phi).
\end{equation}
Thus, by Bayes' theorem,
\begin{align}
    P_{post}(a|x,\Phi) =  \frac{P_{pre}(x|ab,U_\Phi)}{d_A P_{pre}(x|b,U_\Phi)} = \frac{P_{pre}(x|a,\Phi)}{d_A P_{pre}(x|b,U_\Phi)},
\end{align}
which indeed immediately translates to \eqref{postCPTP}. We now have a different perspective on the normalisation factor $f_\Phi(x)$: it quantifies the implicit knowledge about the input ancilla system and how this knowledge affects the postdiction task.  The specification of the quantum channel $\Phi$ contains information about the past of the purifying system system so that postdiction on the quantum channel $\Phi$ is equivalent to postdiction on the purified system but with some added information about the input system.

Indeed, we can arrive at the same formula by using the postdiction probabilities for the purified open system and the simple formula $P(a|b) = P(ab)/P(b)$. Indeed one can verify that
\begin{align}\label{post-channel-ratio}
     P_{post}(a|x,\Phi) = \frac{P_{post}(ab|x,U_\Phi)}{P_{post}(b|x,U_\Phi)} ,
\end{align}
by using \eqref{post-missing-output} and \eqref{post-missing-both}.

While we have used an arbitrary purification, the two equations above hold for any purification of the quantum channel $\Phi$. So, whatever the physically appropriate purification might be, the lesson is the same: the inference asymmetry for a quantum channel derives not from an asymmetry of quantum mechanics, but from an asymmetry in the questions asked.

\subsection{Inference-symmetric channels}
\label{sec:inference-symmetry}

We have seen that not every quantum channel is inference-symmetric. Thanks to the solution \eqref{postCPTP} to the postdiction task in the case of a general quantum channel, we immediately see that a channel $\Phi$ is inference-symmetric if and only if $f_\Phi(x)=1$ for all pure states $x$, namely, if and only if it is identity-preserving:
\begin{equation}
\Phi[\id_A] = \id_X.
\end{equation}
Since quantum channels are trace-preserving, it follows that the input and output spaces are isomorphic. The trace and identity preserving maps are known as the bistochastic quantum maps or unital channels \cite{landau1993birkhoff, mendl2009unital}, and they are the free operations of the resource theory of quantum thermodynamics \cite{chiribella2017microcanonical} with trivial hamiltonians, and the resource theory of purity \cite{streltsov2018maximal}.

Every unitary channel is obviously bistochastic. The noisy operations:
\begin{equation}
    \rho\longmapsto \tr_B U\left[\rho\otimes\frac1{d_B}\id_B\right].
\end{equation}
form a strict subset of the bistochastic channels \cite{chiribella2017microcanonical,ShorTalk}. A noisy operation represents the evolution of a system that undergoes unitary interaction with an ancilla about which nothing is known. These channels are exactly those that are simulated by the tasks considered in the previous section when $Y\equiv B$ and both $Y$ and $B$ are left out of the task.
 
Since the bistochastic channels are the only channels that admit an active time-reversal \cite{chiribella2021wigner}, the equivalence of inference-symmetry and bistochasticity further connects inference-symmetry and time-reversal invariance, as we will see in the next section.

\subsection{More general preparations}
\label{sec:general-preps}

Until now, we limited the preparation to the random element of an orthonormal basis. The reason we considered this is that it reflects the simplest way to interact with a system, namely, to couple to one of its non-degenerate observables. But what if our friend tells us that they prepared one out of $n$, not necessarily orthogonal, states? This situation can also be accounted for using purification.

Say the set of possible initial states states is $\{\psi_i\}_{i=1}^n$, in the $d$-dimensional Hilbert space $A$, let $U$ be the unitary evolution and $\{x_j\}_{j=1}^{d}$ the orthonormal basis of the measurement. Then the prediction probabilities are
\begin{equation}
  P_{pre}(x|\psi_i,U) = |\matrixel{x}{U}{\psi_i}|^2.
\end{equation}
The postdiction probabilities are again found by Bayes' theorem:
  \begin{equation}
  P_{post}(\psi_i|x,U) = \frac{P_{pre}(x|\psi_i,U)P(\psi_i)}{P(x)}.
\end{equation}
Since we are only told the set of possible states, we assume a flat prior over them:
\begin{equation}
  P(\psi_i) = \frac1n,
\end{equation}
and then the probability for the outcome is
\begin{equation}
  P(x) = \sum_{i=1}^{n}\frac1n |\matrixel{x}{U}{\psi_i}|^2 = \tr \left(\ketbra{x}U\left[\rho_{A}\right]\right)
\end{equation}
where we have defined $\rho_A := \sum_i \ketbra{\psi_i}/n$. And thus
  \begin{equation}
  P_{post}(\psi_i|x,U) = \frac{P_{pre}(x|\psi_i,U)}{n \tr \left(\ketbra{x}U\left[\rho_{A}\right]\right)}.
\end{equation}
So the prediction and postdiction probabilities are in general different if the space of possible initial configurations does not represent an orthonormal basis.

We can understand this asymmetry again in terms of implicit knowledge. Let $B$ be a $n$-dimensional Hilbert space, and $\{b_i\}_{i=1}^{n}$ an orthonormal base for it. Choose an orthonormal basis $\{a_i\}_{i=1}^{d}$ on $A$ and find a unitary $U_P$ on  $A\otimes B$ that maps:
\begin{equation}
U_P:\ket{a_1} \otimes \ket{b_i} \longmapsto \ket{\psi_{i}} \otimes \ket{b_i}.
\end{equation}
Also define the unitary $U': A\otimes B \rightarrow X \otimes B$ given by
\begin{equation}
U' = (U \otimes \id_B)    \circ U_P.
\end{equation}
We can relate the prediction and postdiction probabilities for these two games. For prediction we have:
\begin{equation}
  P_{pre}(x|\psi_i,U) = P_{pre}(x|a_1b_i,U')
\end{equation}
since, for an arbitrary basis $\{y_i\}_{i=1}^n$ of $B$:
\begin{equation}
\begin{aligned}
  P_{pre}(x|a_1b_i,U') &= \sum_{j=1}^n \matrixel{xy_j}{U'}{a_1b_i} \\
  &= \sum_{j=1}^n |\matrixel{xy_j}{U\otimes\id_B}{\psi_ib_i}|^2 \\
  &= |\matrixel{x}{U}{\psi_i}|^2\cdot\sum_{j=1}^n |\braket{y_j}{b_i}|^2\\
 P_{pre}(x|a_1b_i,U') &=    P_{pre}(x|\psi_i,U),
  \end{aligned}
\end{equation}
where we have used the definitions of $U'$ and $U_P$ to obtain the second equality.

For postdiction we have
\begin{equation}\label{post-general-preparation}
  P_{post}(\psi_i|x,U) =\frac{P_{post}(a_1b_i|x,U')}{P_{post}(a_1|x,U')},
\end{equation}
which is entirely analogous with \eqref{post-channel-ratio}.
The proof is just a matter of expressing nominator and denominator in terms of probabilities for the original task:
\begin{equation}
  \begin{aligned}
      P_{post}(a_1b_i|x,U') &=\frac{1}{n} \sum_{j=1}^n P_{post}(a_1b_i|xy_j,U') \\
      &=\frac{1}{n} \sum_{j=1}^n P_{pre}(xy_j|a_1b_i,U') \\
      &=\frac{1}{n} P_{pre}(x|a_1b_i,U') \\
      P_{post}(a_1b_i|x,U') &= \frac{1}{n} P_{pre}(x|\psi_i,U)
  \end{aligned}
\end{equation}
and so
\begin{equation}
    \begin{aligned}
        P_{post}(a_1|x,U') &= \sum_{i=1}^{n}  P_{post}(a_1b_i|x,U')\\
      &= \sum_{i=1}^{n} \frac1n  P_{pre}(x|\psi_i,U)  \\ 
     P_{post}(a_1|x,U')   &=\tr \left(\ketbra x  U[\rho_A]\right). 
    \end{aligned}
\end{equation}

Thus, cases where we have more general preparations can be understood in terms of preparations on orthonormal bases. From \eqref{post-general-preparation}, in analogy with \eqref{post-channel-ratio}, that the preparation of non-orthonormal states contains some implicit information about an ancilla system that allows to prepare the non-orthogonal states. 
Again, we see that in this purified game, the only way to achieve postdiction probability 1 is to marginalise over $b_i$, \textit{i.e.} not guess at all.

This method can be further generalised to the preparation of an arbitrary set of density operators by adding a purifying system that gets traced over in the prediction task, and is assigned a flat prior in the postdiction task. The asymmetry again can be understood as an asymmetry in the assumed knowledge in the tasks.

\section{Relation between time-reversal and postdiction}
\label{sec:timereversal}

We have explored the symmetry between guessing the future and guessing the past. We have seen that in the case of a closed system, the past is quantitatively as uncertain as the future, in the sense that the Born rule can also be used, without modification, to guess a past event based on information about a future event. In the case of a general quantum channel, this symmetry is broken: the prediction and postdiction probabilities are given by different formulas. However, the reason for this is that the implementation of a general quantum channel requires the preparation of a system in a given state, so that knowing a channel was implemented confers information about the input system, breaking the symmetry between prediction and postdiction. 
Quantum probabilities by themselves have no regard for the direction of time.

In this section, we examine another way in which the quantum probabilities do not distinguish between past and future by examining the more familiar notion of time symmetry: that of time-reversal symmetry. There are two notions of time-reversal, which we could call active and passive, or operational and descriptive. These notions of time-reversal are slightly different from, but closely related to, those that would be applied to a system of differential equations, where one maps solutions to solutions.

In the passive time-reversal of a system, one simply examines the changes in the system in the reverse order, starting from the future and moving towards the past. Applied to the inference tasks we have been considering, passive time-reversal amounts to switching from the prediction task to the postdiction task and \textit{vice versa}. The previous sections have thus been an investigation of passive time-reversal symmetry; we proved that passive time-reversal symmetry is equivalent to inference-symmetry.

Active time-reversal consists instead in finding a process that undoes the change that was brought by a previous transformation. In the context of the inference tasks, active time-reversal consists in considering a new, time-reversed task.

\subsection{Time-reversed task with unitary channel}
\label{sec:time-reversed-unitary}

Unitary maps are invertible, and thus for every evolution $U$ of a closed quantum system, there exists a time-reversed evolution given by the adjoint $U^\dagger$. 

\begin{mydef}[Time-reversed task]
Consider a task in which a closed system is prepared in a basis $\{a_i\}_{i=1}^d$ and measured in a basis $\{x_i\}_{i=1}^d$ after undergoing the unitary evolution $U$. In the time-reversed task, the system is prepared in  $\{x_i\}_{i=1}^d$ and measured in $\{a_i\}_{i=1}^d$ after undergoing the evolution $U^\dagger$.
\end{mydef}
\noindent It follows immediately from the properties of the inner product that
\begin{equation}
\label{time-rev-closed}
P_{pre}(a|x,U^\dagger) = P_{pre}(x|a,U).
\end{equation}
While this is result is trivial to derive, it is nevertheless profound, as it compounds with the inference symmetry
\begin{equation}\label{inferenceU}
P_{pre}(x|a,U) = P_{post}(a|x,U),
\end{equation}
to show how little regard the probabilities of closed quantum systems have about the direction of time.
Indeed, given two pure states $a$ and $x$ of the corresponding Hilbert spaces, the \textit{same} quantity
\begin{equation}\label{inner}
|\matrixel{x}{U}{a}|^2
\end{equation}
is the solution for four conceptually distinct taks:
\begin{itemize}
\item $P_{pre}(x|a,U)$, for the prediction task with unitary evolution $U$,
\item $P_{post}(a|x,U)$, for the postdiction task with unitary evolution $U$,
\item  $P_{pre}(a|x,U^\dagger)$, for the prediction task with time-reversed unitary evolution $U^\dagger$, and
\item $P_{post}(x|a,U^\dagger)$, for the postdiction task with time-reversed unitary evolution $U^\dagger$.
\end{itemize}
The familiar Born rule, which is normally thought to apply only to the first case, actually applies to all four of these cases. In each one of them, $x$ can be either the future or the past event, and can either be the known or unknown in the scenario.

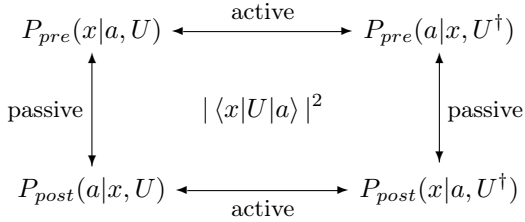
\begin{diagram}
P_{pre}(x|a,U) & \rEquiv^{\mbox{\small active}} & P_{pre}(a|x,U^\dagger) \\
\dEquiv^{\mbox{\small passive}} &~~~~ |\matrixel{x}{U}{a}|^2~~~~ & \dEquiv_{\mbox{\small passive}} \\
P_{post}(a|x,U) & \rEquiv_{\mbox{\small active}} & P_{post}(x|a,U^\dagger)
\end{diagram}

Let us now consider the case of open systems and study the time-reversed version of the tasks examined in section \ref{sec:open}. If we neglect the outcome of the measurement on $B$, we can write:
\begin{align}
	P_{pre}(a|xy,U^\dagger) 
	&= \sum_{i=1}^{d_B}P_{pre}(ab_i|xy,U^\dagger) \\
	&= \sum_{i=1}^{d_B}P_{pre}(xy|ab_i,U) \\
	P_{pre}(a|xy,U^\dagger) 
	&= P_{post}(a|xy,U) \label{pre-a-xy},
\end{align}
where we have used the time-reversal symmetry \eqref{time-rev-closed} in the second equality, and \eqref{post-missing-output} in the third.
This equation relates the probability of the prediction task to the probability of the time-reversed postdiction task. In both cases, we are calculating the probability of event $a$ based on knowledge of event $xy$. However, in one case $xy$ is to the past of $a$ and in the other case the opposite is true.
In a similar fashion, we can derive the following equations:
\begin{align}
	P_{pre}(ab|x,U^\dagger)  &= P_{post}(ab|x,U), \label{pre-ab-x}\\
	P_{post}(xy|a,U^\dagger)  &= P_{pre}(xy|a,U), \label{post-xy-a}\\
	P_{post}(x|ab,U^\dagger)  &= P_{pre}(x|ab,U), \label{post-x-ab}
\end{align}
for when only one side is being ignored, as well as
\begin{align}
	P_{pre}(a|x,U^\dagger)  &= P_{post}(a|x,U), \label{pre-a-x}\\
	P_{post}(x|a,U^\dagger)  &= P_{pre}(x|a,U), \label{post-x-a}
\end{align}
for when data is being ignored on both sides.
Recall the discussion in section \ref{sec:normalisation} about the normalisation of the identity that is required to calculate the quantities on the right hand side. The same applies here, which reinforces the lesson: the direction of time is irrelevant to the normalisation of the identity. What matters is the direction of inference:

\begin{equation}
  P_{pre}(x|a,U) = \begin{tikzpicture}
	\begin{pgfonlayer}{nodelayer}
		\node [style=bigtrapezium] (0) at (0, 0) {$U$};
		\node [style=effect] (1) at (-0.625, 1.5) {$x$};
		\node [style=state] (4) at (-0.625, -1.5) {$a$};
		\node [style=discard] (5) at (0.625, 1.275) {};
		\node [style=identity, label={right:$\;\frac{1}{\,d_{\!B}}$}] (6) at (0.625, -1.28) {};
	\end{pgfonlayer}
	\begin{pgfonlayer}{edgelayer}
		\draw [style=black] (1) to (4);
		\draw [style=black] (5) to (6);
	\end{pgfonlayer}
\end{tikzpicture} \!\!\!\!= \begin{tikzpicture}
	\begin{pgfonlayer}{nodelayer}
		\node [style=daggerbigtrapezium] (0) at (0, 0) {$U$};
		\node [style=effect] (1) at (-0.625, 1.5) {$a$};
		\node [style=state] (4) at (-0.625, -1.5) {$x$};
		\node [style=discard, label={right:$\;\frac{1}{\,d_{B}}$}] (5) at (0.625, 1.275) {};
		\node [style=identity] (6) at (0.625, -1.28) {};
	\end{pgfonlayer}
	\begin{pgfonlayer}{edgelayer}
		\draw [style=black] (1) to (4);
		\draw [style=black] (5) to (6);
	\end{pgfonlayer}
\end{tikzpicture} \!\!\!\!\!=   P_{post}(x|a,U^\dagger).
\end{equation}

The equation above\footnote{In the diagrammatic calculus, flipping a diagram along a horizontal axis denotes taking the hermitian conjugate.} refers to inference tasks in which we have information about the system $A$ and want to guess something about the system $X$. It does not matter if $A$ is in the past or the future of $X$. The answer is the same. There is no way of telling which way the arrow of time is pointing.

\subsection{Time-reversed task with quantum channel}
\label{sec:time-reversed-quantum-channel}

The natural candidate for the time-reversed task is the one with the adjoint $\Phi^\dagger$ of the quantum channel $\Phi$. In fact, Chiribella \textit{et. al.} proved that this is essentially the unique way of defining the active time-reversal of a quantum channel \cite{chiribella2021wigner,chiribella2020quantum}. We must distinguish two cases, as the adjoint of a CPTP map may or may not be a CPTP map.

\medskip

The adjoint $\Phi^\dagger$ is trace-preserving if and only if $\Phi$ is unit preserving since, by the definition of the adjoint, for all $\rho\in\mathcal{L}{(X)}$:
\begin{equation}
    \tr \Phi^\dagger[\rho] = \tr \id_A\circ\Phi^\dagger[\rho] = \tr \Phi[\id_A]\circ\rho.
\end{equation}
Furthermore, we have seen that the bistochastic channels are exactly the inference symmetric channels. In analogy with the discussion above on the closed systems, for a bistochastic channel $\Phi$, the quantity
\begin{equation}
	\tr \ketbra{x} \circ \Phi[\ketbra{a}],
\end{equation}
given by the generalised Born rule, yields the numerical value of four \textit{a priori} conceptually distinct quantities:\begin{itemize}
\item $P_{pre}(x|a,\Phi)$, for  the prediction task with bistochastic channel $\Phi$,
\item $P_{post}(a|x,\Phi)$, for the postdiction task with bistochastic channel $\Phi$,
\item  $P_{pre}(a|x,\Phi^\dagger)$, for the prediction task with time-reversed bistochastic channel $\Phi^\dagger$ and
\item $P_{post}(x|a,\Phi^\dagger)$, for the postdiction task with time-reversed bistochastic channel $\Phi^\dagger$.
\end{itemize}
Thus postdiction-symmetry is intimately linked with time-reversal invariance even in the context of open quantum systems. 

In general, however, the adjoint of a CPTP map might fail to be trace non-increasing. So it might not only fail to represent a quantum channel, but may not even be a quantum map, \ie part of an operation. In this case, an active time-reversal of the corresponding quantum channel does not exist, as has been recently shown in \cite{chiribella2021wigner,chiribella2020quantum}.

\subsection{Quantum channels towards the past}

A quantum channel might not have an active time-reversed version. Nevertheless, we can learn something striking by looking at the time-reversed task of a \textit{purification} of this quantum channel: CPTP maps can describe \textit{postdictions} in situations in which part of the future data is left implicit and fixed. Consider the time-reversed version of the purified task in section \ref{sec:purification}. The system is prepared in some state corresponding to the basis $\{x_iy_j\}$ and measured on some basis $\{a_ib_j\}$ after undergoing the transformation described by ${U_\Phi}^{\!\dagger}$. Using the time-symmetry of unitary open systems \eqref{post-x-ab}, observe that
\begin{equation}
	P_{pre}(x|a,\Phi) = P_{pre}(x|ab,U_\Phi) =   P_{post}(x|ab,U_\Phi^\dagger).
\end{equation}
That is, the quantity
\begin{equation}\label{CPTPBorn}
  \tr \ketbra{x}\Phi[\ketbra{a}] 
\end{equation}
is the solution to two different tasks:
\begin{itemize}
\item $P_{pre}(x|ab,U_\Phi)$ for the prediction task with unitary $U_\Phi$, and
\item $P_{post}(x|ab,U^\dagger_\Phi)$ for the time-reversed postdiction task with $U_\Phi^\dagger$.
\end{itemize}
Thus one can take the quantity \eqref{CPTPBorn} to relate to an inference towards the past, in a situation in which part of the future events are only implicitly described.
This furthers the argument that the inference-asymmetry of the CPTP maps is not an asymmetry related to an intrinsic direction of time in the details of a quantum process, but in an asymmetry in the data about a system. The input of a CPTP map does not necessarily lie to the past of the output, it can also lie to its future.

This last insight will play a major role in dispelling a source of confusion regarding the time-orientation of quantum phenomena.

\section{The arrow of inference}
\label{sec:causality}
In this section, we discuss two closely related asymmetric aspects of the operational formalism, encapsulated by two maxims: ``there exists a unique deterministic effect'' and ``no signalling from the future''. These expressions reflect mathematical properties of the theory that are often taken as evidence of an asymmetry between the past and the future.  Here we show that these properties reflect the asymmetry due to the arrow of inference, which is in principle independent of the arrow of time.

\subsection{``There exists a unique deterministic effect''}
\label{sec:deterministic-effect}

As a mathematical statement, given the definitions, this maxim is correct. A quantum map is deemed \textit{deterministic}, or \textit{causal}, if it is trace preserving. An \textit{effect} is a quantum map from a Hilbert space to the trivial Hilbert space. Thus a \textit{deterministic effect} is an effect that is trace-preserving. It is easy to see that the only such effect is the one that maps a state to its trace, \textit{i.e.} the discard operator. In this precise sense the discard is the only deterministic effect.
 
As a statement of the irreducible uncertainty of quantum phenomena, the maxim is also correct: the only way to be certain about a prediction in all cases is not to guess at all. In the context of prediction tasks, discarding means not trying to predict anything about the system and is represented by the identity operator as an effect (see section \ref{sec:normalisation}). Not guessing is the only way to always guess right, and this is the physical content of ``there exists a unique deterministic effect.''

The uniqueness of the deterministic effect is sometimes understood as the conservation of probability, a mere aspect of inference (see for example \cite{barrett2007information,schmid2020unscrambling}). However, the maxim is sometimes taken to signify something fundamental about the distinction between the past and the future \cite{oreshkov2015operational,coecke2017timereverse}, and this is not correct. 
In fact, we can define by analogy what it means to discard something in the past. If discarding in the future means marginalising prediction probabilities, then discarding in the past means marginalising postdiction probabilities. Discarding in the past is also represented by the identity operator when computing probabilities---this time in the \textit{input} side, as we have seen at the end of section \ref{sec:open}. Indeed, looking at the formula \eqref{postCPTP} for postdiction with a general quantum channel, we see that the only way to have probability $1$ is to discard the system, \ie by not guessing at all.

Thus, the uniqueness of the discard operator is not a consequence of the time-orientation of quantum phenomena. It is instead a manifestation of thoroughly time-symmetric quantum indeterminism in the context of prediction.

\subsection{``No signalling from the future''}
\label{sec:no-signalling-from-the-future}

This maxim means that the probabilities of the outcomes of a quantum operation do not depend on the nature of a later operation.  To lay down some notation, we reproduce the standard proof of this property, which is an immediate consequence of the equations \eqref{completeness} and \eqref{sequential-prob}. In the next subsection we comment on the physical reasons for this and show that it too is in fact a property of the arrow of inference, not a property of time.

Suppose a system starts in a state $\rho$, and the operation $\mathcal{E}^{A\rightarrow D}=\{E_x\}$ is applied. The quantity
\begin{equation}
P_{pre}(x|\rho, \mathcal{E}) = \tr E_x[\rho]
\end{equation}
is the probability of observing outcome $x$.

Suppose instead that the operation $\mathcal{E}$ is immediately followed by another operation $\mathcal{F}^{D\rightarrow Z}=\{F_y\}$. The probability for the outcomes $x$ and $y$ is given definition by~\eqref{sequential-prob}:
\begin{equation}
	P_{pre}(xy|\rho,\mathcal{F}\circ\mathcal{E}) = \tr F_y\left[E_x[\rho]\right]
\end{equation}
An immediate application of probability theory and the completeness equation \eqref{completeness} yields
\begin{equation}
  P_{pre}(x|\rho,\mathcal{F}\circ\mathcal{E}) = \sum_y \tr F_y\left[E_x[\rho]\right] = \tr E_x[\rho], 
\end{equation}
and we thus have the identity
\begin{equation}\label{no-signalling-future}
  P_{pre}(x|\rho,\mathcal{F}\circ\mathcal{E}) = P_{pre}(x|\rho, \mathcal{E}).
\end{equation}
The probabilities of the outcomes of the first operation are independent on the nature of the second operation.\footnotemark\; The case in which the operation $\mathcal{F}$ is performed and the case in which it is not are indistinguishable by looking only at what happens at $\mathcal{E}$. 
A future experimenter cannot affect the statistics in the present by manipulating a system in the future, hence ``no signalling from the future.''

What about the opposite direction? By linearity and \eqref{unseen-operation} we have
\begin{equation}
  P_{pre}(y|\rho,\mathcal{F}\circ\mathcal{E}) =  P_{pre}(y|\mathcal{E}[\rho],\mathcal{F}),
\end{equation}
so that the probability of $y$ clearly depends on the first operation $\mathcal{E}$.

\foottext{Note that the no-signalling from the future property \eqref{no-signalling-future} can also be seen as a motivation for the definition \eqref{sequential-prob} for the probabilities of the outcomes of sequential operations in the first place.
Indeed by linearity, these can be rewritten as:
\begin{equation}
\begin{aligned}
	\tr F_y\left[E_x[\rho]\right]
	&= \tr \left[F_y\left(\frac{E_x[\rho]}{\tr E_x[\rho]}\right)\right]\cdot\tr E_x[\rho] \\
\end{aligned}
\end{equation}
so that, by setting $\rho_x = {E_x[\rho]}/{\tr E_x[\rho]}$ one can write:
\begin{equation}
	P_{pre}(xy|\rho,\mathcal{F}\circ\mathcal{E})
	= P_{pre}(y|\rho_x,\mathcal{F})\cdot P(x|\rho,\mathcal{E}).
\end{equation}}

Thus the future operation does not affect the probabilities of past events, while the past operation affects the probability of future events. One seems compelled to conclude that this reveals a time-asymmetry in quantum theory. But this conclusion is too quick. In previous sections, we have encountered various situations in which asymmetric aspects of quantum theory should not be ascribed to a difference between past and future, but to the directionality of inference and an asymmetry of the data. We have also seen that the operational formulations assume that the arrow of time and the arrow of inference point the same way. We should be cautious.

\subsection{``No signalling from the further unknown''}
\label{sec:further-unknown}


The calculation above in fact shows that the future operation does not affect the \textit{prediction} probabilities of events in its past, while the past operation affects the \textit{prediction} probabilities of events in its future. 
In other words, if somebody in the past of both operations is trying to guess what would be the outcome of the first operation, they can safely discard any information about the second operation that might be available at that time.\footnotemark\;  
Put another way: an event further away from the data does not affect the prediction probabilities closer to the data.

\foottext{To be sure, having knowledge of the outcome of the second operation \textit{does} provide an advantage in guessing the outcome of the first one, since 
\begin{equation}
  P(x|y,\rho,\mathcal{F}\circ\mathcal{E}) = {\tr F_y[E_x[\rho]]}/{\tr F_y[\mathcal{E}[\rho]]}.
\end{equation}
  This data however is unavailable to somebody sitting in the past of both operations.
}

A similar statement holds also when the arrow of inference points toward decreasing time, \ie when doing postdiction. As we have already seen in section \ref{sec:time-reversed-quantum-channel}, one can understand a CPTP map as a shorthand to calculate a postdiction probability when a future event is implicitly considered fixed. The same is generally true of \textit{any} operation: it can serve as a shorthand to aid the calculation of postdiction probabilities. Then \eqref{no-signalling-future}, understood as a statement about postdiction probabilities, tells us that what happened in the further past does not affect our ability to infer what happens in the closer past. When predicting, an event does not affect prediction probabilities about an event in its past. When postdicting, an event does not affect postdiction probabilities of events in its future. In both cases, there is no signalling from the further unknown.  

As an illustration, let us consider a purification of the two operations $\mathcal{E}^{A\rightarrow D}$ and $\mathcal{F}^{D\rightarrow Z}$ above. Let us assume for simplicity that there is no need to discard a part of the output system (by handling the more general case we would reach the same conclusions but with more typographical effort). Then there exists Hilbert spaces $B,C, X$ and $Y$, unitaries $U_\mathcal{E}:{A\otimes B\rightarrow X\otimes D}$ and $U_\mathcal{F}:{D\otimes C\rightarrow Y\otimes Z}$, pure states $\ket b\in B$ and $\ket c \in C$ as well as orthonormal bases for $X$ and $Y$ labelled by $x$ and $y$ respectively, such that for all $\rho\in\mathcal{L}(A)$ and $\sigma\in\mathcal{L}(D)$:
\begin{align}
    E_x[\rho] &= \tr_X \ketbra{x}\circ U_\mathcal{E}[\rho\otimes \ketbra{b}], \\
    F_y[\sigma] &= \tr_Y \ketbra{y}\circ U_\mathcal{F}[\sigma\otimes \ketbra{c}],
\end{align}
meaning that:
\begin{equation}\label{tikz-povms}
\scalebox{1.2}{\begin{tikzpicture}
	\begin{pgfonlayer}{nodelayer}
		\node [style=state] (0) at (2.725, -2.375) {$b$};
		\node [style=none] (1) at (1.35, -2.875) {};
		\node [style=state] (2) at (4.1, -2.35) {$c$};
		\node [style=bigtrapezium] (3) at (2, -0.75) {$U_\mathcal{E}$};
		\node [style=bigtrapezium] (4) at (3.4, 0.875) {$U_\mathcal{F}$};
		\node [style=effect] (5) at (2.725, 2.4) {$y$};
		\node [style=effect] (6) at (1.35, 2.375) {$x$};
		\node [style=none] (7) at (4.1, 2.875) {};
		\node [style=none] (8) at (0, 0) {$=$};
		\node [style=none] (9) at (-2, -2.875) {};
		\node [style=none] (10) at (-2, 2.875) {};
		\node [style=trapezium] (11) at (-2, -1.125) {$E_x$};
		\node [style=trapezium] (12) at (-2, 1) {$F_y$};
	\end{pgfonlayer}
	\begin{pgfonlayer}{edgelayer}
		\draw [style=black] (6) to (1.center);
		\draw [style=black] (5) to (0);
		\draw [style=black] (7.center) to (2);
		\draw [style=black] (10.center) to (9.center);
	\end{pgfonlayer}
\end{tikzpicture}}.
\end{equation}
We can write down the prediction probabilities for the purified task in terms of the operations:
\begin{equation}
P_{pre}(xy|abc, U_\mathcal{F}U_\mathcal{E}) =   \tr F_y[E_x[a]].
\end{equation}
Now we can use the property \eqref{post-x-ab} of time-reversed unitary tasks to also write
\begin{equation}
P_{post}(xy|abc, U_\mathcal{E}^\dagger U_\mathcal{F}^\dagger) = \tr F_y[E_x[a]].
\end{equation}

Thus the quantities $\tr F_y[E_x[a]]$ are also postdiction probabilities for a different scenario, in which \textit{future} events $b$ and $c$ are held fixed, and this knowledge is used to guess something happening to their past ($x$ and $y$):
\begin{equation}
  \scalebox{1.2}{\begin{tikzpicture}
	\begin{pgfonlayer}{nodelayer}
		\node [style=effect] (0) at (2.725, 2.375) {$b$};
		\node [style=effect] (2) at (4.1, 2.35) {$c$};
		\node [style=daggerbigtrapezium] (3) at (2, 0.75) {$U_\mathcal{E}$};
		\node [style=daggerbigtrapezium] (4) at (3.4, -0.875) {$U_\mathcal{F}$};
		\node [style=state] (5) at (2.725, -2.4) {$y$};
		\node [style=state] (6) at (1.35, -2.375) {$x$};
		\node [style=identity] (7) at (4.1, -2.15) {};
		\node [style=none] (8) at (0, 0) {$=$};
		\node [style=trapezium] (11) at (-1.6, -0.875) {$E_x$};
		\node [style=trapezium] (12) at (-1.6, 0.75) {$F_y$};
		\node [style=state] (13) at (-1.6, -2.375) {$a$};
		\node [style=discard] (14) at (-1.6, 2.125) {};
		\node [style=effect] (15) at (1.35, 2.35) {$a$};
	\end{pgfonlayer}
	\begin{pgfonlayer}{edgelayer}
		\draw [style=black] (5) to (0);
		\draw [style=black] (7) to (2);
		\draw [style=black] (14) to (13);
		\draw [style=black] (15) to (6);
	\end{pgfonlayer}
\end{tikzpicture}}.
\end{equation}
In this case, the operation $\mathcal{F}$  contains information about something that happens \textit{earlier}
 in the system, namely, the interaction $U_\mathcal{F}^\dagger$ between the subsystems $Y$ and $Z$, and the outcome of the measurement of $C$. All of this information is irrelevant when postdicting only the preparation on $X$:
\begin{equation}
  \sum_y P_{post}(xy|abc, U_\mathcal{E}^\dagger U_\mathcal{F}^\dagger) = P_{post}(x|abc, U_\mathcal{E}^\dagger).
\end{equation}
Thus, when postdicting, the outcome and nature of the earlier operation is irrelevant to the probabilities for the outcome of the later operation.

\subsection{Why we \textit{can} signal from the past}
\label{sec:agents}

We have seen that the operation $\mathcal{E}$ affects the prediction probabilities for the outcome of $\mathcal{F}$. How is this related to the notion of \textit{signalling}? Let's imagine two parties, one located where $\mathcal{E}$ takes place and one where $\mathcal{F}$ takes place. Let's call them Eve and Fred. How can Eve send a signal to Fred? 
The barebone scenario for signalling to the future is that Eve's operation is simply a state preparation of a qubit, and Fred's is a projective measurement. Eve, who knows the basis on which Fred will measure, \textit{chooses} one of the two states so that she fully determines the outcome of Fred's operation. Like this, she can send one bit of information. 

Notice, however, that Eve's ability to choose is crucial to this protocol.
If she cannot pick what state to send to Fred, she cannot send a message to him. All she can do is try to predict what will be the outcome of Fred's measurement, once---and if---she knows what state she sent him. How does somebody prepare a system in a specific state? In practice, this is done by subjecting the system to a maximal test and discarding the systems yielding unwanted results \cite{Peres}. Alternatively, one can apply a unitary transformation to the system, conditional on the outcome of the maximal test. Both of these procedures require an increase in the entropy of the universe, as the first involves picking and choosing \cite{rovelli2020agency, milburn2020physical} and the second is an instance of erasure so Landauer's principle applies \cite{landauer1961irreversibility}, a rigorous proof of which just appeared \cite{myrvold2020shakin}. Thus, signalling is a concept beyond single quantum transition probabilities and has its origins in the thermodynamic arrow of time.

We are time oriented creatures, we know more about the past than about the future, we mostly try to guess the future.  Quantum probabilities do not care if you are making guesses about the future or the past. They are about predicting what is unknown from what is known. The difference between what is known and what is unknown is at the origin of the time asymmetric maxims of the operational formulations.  The first maxim arises from the fact that postdiction scenarios are rarer than prediction ones in practice. If we want to learn about the past, we find there are plenty of records about past events in the present. The existence of traces is not a property of the probabilities of individual quantum systems, so the fact that we rarely have to guess about the past the same way we have to guess about the future is no evidence of a time asymmetry of the physics of quanta. 

The grip of the second maxim on the community is more subtle. It rests on the notion of signalling, which is itself tightly linked with ideas of causation and agency, concepts we have strong intuitions for and rely on daily. We humans make choices and these choices influence our future (not our past). The same goes for a lot of systems around us: when my laptop suddenly ``decides'' to break, it will affect my ability to finish a future paper (not a past one). Is this time-orientation a direct consequence of some time asymmetry in quantum phenomena? Hopefully, by now, the answer is clear. Is causation a fundamental property of the world in some other way? This question has received surprisingly little attention from the physics community at large \cite{price1997time}. This question needs to be addressed carefully if one hopes of extending these operational formalisms to probe physics outside laboratories. The notion of causes always preceding effects is strongly related to notions of agency. And agency is a perspectival property, stemming from a partial description of systems and the presence of an entropy gradient \cite{rovelli2020agency, milburn2020physical}.  One should be cautious in extending notions of time oriented causation all the way to elementary physics. 

\section{Time orientation of other formalisms}

Before concluding the paper, we add some comments regarding the time orientation in other formalisms. 

In the de Broglie-Bohm interpretation~\cite{bohm1952suggested,goldstein2017bohmian}, the particle and the pilot-wave obey time-reversal invariant dynamics. The Everettian wavefunction \cite{saunders2010many, vaidman2018manyworlds} also obeys time-reversal invariant dynamics. Branching towards the future is interpreted as a past low entropy condition: systems were uncorrelated in the past, hence the vanishing of von-Neumann relative entropy. Alternatively, it can be interpreted as a perspectival aspect of the Everettian relative state determined by an observation in the present. 

The Copenhagen quantum state is correlated with past observations, not future ones, but the state is, of course, unobservable and its ontological status is debated. Its empirical content is given by the probabilities it allows to calculate. If we ask time-symmetric questions, the probabilities we obtain are time-symmetric \cite{aharonov1964time}. The state is assumed to be correlated with past events because we are using past data to infer about the future. If we want to guess about the past, we could just as well use a quantum state correlated with a future event \cite{watanabe1955symmetry,speirits2017retrodiction,rovelli2016argument}.

In QBism \cite{Fuchs2019}, quantum theory is interpreted as a means to aid decision making, allowing an agent to calculate the probabilities of the consequences of their interactions with the world.  Because decision-making agents play a central role in bringing about the world according to the QBists \cite{fuchs2016participatory}, the ontology of the theory is fundamentally time oriented.

Lab measurements generally involve decoherence and amplification of a microscopic phenomenon to the macroscopic realm, both of which rely on the entropy gradient. Views of quantum theory that insist that the only real events are of this kind are therefore time-oriented, even though the probabilities themselves might be time-symmetric. These views also imply that there is no interpretation of the theory outside of the macroscopic approximation.

According to the relational interpretation of quantum mechanics (RQM) \cite{rovelli1996relational, rovelliSpaceBlueBirds2018}, facts happen at every interaction between any two systems, but the facts are relative to the systems involved in the interaction. The quantum state only plays a computational role in this interpretation.  Decoherence comes into play to `stabilise' a fact \cite{dibiagio2020stable}, so that one might ignore its relational nature, which is manifest in interference effects. Decoherence requires information loss and an increase in entropy. Hence RQM is a time-symmetric formulation of quantum theory, but the dynamics of \emph{relative} facts is  time symmetric while the dynamics of \emph{stable} facts is time oriented.  

 

\section*{Conclusion}

Quantum theory is not about predicting the future, it is about time-symmetric conditional probabilities relating events.
The directionality internal to the theory is the arrow of inference, the difference between known and unknown.

There are \emph{formulations} of quantum theory that break time reversal symmetry and use time oriented theoretical notions. These either refer to non-observable entities, or to assumptions about the time orientation of inference problems, or to the entropic time orientation of decoherence. Examples of such formulations are provided by the use of a quantum state determined by \emph{past} interactions in the Copenhagen interpretation, and the use of the quantum operations described above. The time orientation of operations is due to them being high-level notions with a built-in assumption about time asymmetric capabilities of the experimentalists. 

The time orientation of the formalisms we use is determined by the  common boundary conditions we set for the physical processes we study: they come from the assumptions about the agent interacting with the system and the conditions she imposes on it.  The agent is not directly modelled in the theory and is instead represented by the inferential boundary conditions and choice of operations, the exogenous variables of \cite{myrvold2020science}.  Time orientation is in this way external to the elementary quantum process being modelled. It can, in principle, be entirely accounted for at the level of statistical mechanics as a consequence of the existence of an entropy gradient, namely past low entropy \cite{mlodinow2014relation,rovelli2020memory, rovelli2020agency,milburn2020physical}.

Some authors \cite{oreshkov2015operational,aharonov1964time} have built time-symmetric theories to replace quantum theory. 
However, the results of sections \ref{sec:unitary} and \ref{sec:timereversal} show that quantum theory is already time-symmetric as it is. The transition probabilities calculated with quantum theory are blind to the direction of time. 
The probabilities of closed quantum systems are inference symmetric and time-reversal invariant. Thus, when accounting for all the relevant degrees of freedom, the predictions of quantum theory are thoroughly time-agnostic. The probabilities of open quantum systems are in general neither inference symmetric nor time-reversal invariant. We have shown that the asymmetry between prediction and postdiction in this case is only a consequence of treating the two problems asymmetrically, by assuming more knowledge in one case than in the other. Both inference asymmetry and the failure of time-reversal invariance of quantum channels can be understood in the same terms. This asymmetry\footnotemark\  is not intrinsic to the mechanical theory, but is, rather, an asymmetry of the questions we humans ask using the theory. 

\foottext{This situation is reminiscent of the one in classical statistical mechanics, in which one can prove that entropy increases \textit{both} towards the past \textit{and} towards the future of the given initial state \cite{price1997time,myrvold2020science}. In both situations we use time-symmetrical mechanical laws to extrapolate from a state of limited knowledge to a state at other times. This results in even more limited knowledge. The fact that entropy was lower in the past, is a fact external to the laws of mechanics. And the fact that we are mostly interested in extrapolating towards the future is in turn a consequence of the low entropy in the past.}

\begin{acknowledgments}
The authors thank Giulio Chiribella and his QICI group, Lucien Hardy, Laurent Freidel, Flaminia Giacomini, Francesca Vidotto, and especially Marios Christodoulou for critical and helpful discussions on early versions on this work. The authors also thank the three anonymous referees for thorough and helpful critical feedback.

This work was made possible through the support of 
the  FQXi  Grant  FQXi-RFP-1818 and of 
the ID\# 61466 grant from the John Templeton Foundation, as part of the ``The Quantum Information Structure of Spacetime (QISS)'' Project (\href{qiss.fr}{qiss.fr}). 
The work of P.D. is partially supported by the grant 2018-190485 (5881) of the Foundational Questions Institute and the Fetzer Franklin Fund.

The diagrammatic formulas were realised with TikZit~(\href{https://tikzit.github.io/}{tikzit.github.io}).
\end{acknowledgments}
\vfill

\bibliography{short}

\end{document}